# Generating eukaryotic reference genome assemblies: Earth BioGenome Project quality standards and recommendations

## Authors

Kerstin Howe*1, Dolores R Agius2, Tyler Alioto3,4, Jean-Marc Aury5, Roberto Biello6, Mark Blaxter1, Matthew W. Brown7, Tom Brown1, Bernat Burriel-Carranza8, Haoyu Cheng9, Dario Copetti10, Solenne Correard11, Fernando Cruz3,4, Paul Davis1, Aleksandra Denisova12,13, Giulio Formenti12, Pedro A F Galante14, Caroline Howard1, Kamil S. Jaron1, Erich D. Jarvis12, Sergei Kliver6, Sergey Koren15, Ilia J Leitch16, Harris Lewin17, Heng Li18, Kerstin Lindblad-Toh19,20, Jack A. Medico12, Matthieu Muffato1, Zemin Ning1, Javan Okendo15, Shujun Ou21, Marc Palmada-Flores22, Brendan J. Pinto15, Aurora Ruiz-Herrera23, Abdoallah Sharaf24, Ying Sims1, Jeramiah J Smith25, Marco Sollitto 12,26, Cibele G. Sotero-Caio1, Marcela Uliano-Silva1,27, Sibelle T. Vilaça28, Robert M. Waterhouse29, Melissa A. Wilson15, Chentao Yang30, Guojie Zhang31, Chenxi Zhou32, Yang Zhou33, Richard Durbin32

1Wellcome Sanger Institute, Wellcome Genome Campus, Hinxton, Cambridge CB10 1SA. UK
2Botany and Mycology Unit Finnish Museum of Natural History & Department of Organismal and Evolutionary Biology Faculty of Biological and Environmental Sciences University of Helsinki Finland
3Centro Nacional de Análisis Genómico (CNAG), Baldiri Reixac 4, 08028 Barcelona, Spain
4Universitat de Barcelona (UB), 08028 Barcelona, Spain
5Génomique Métabolique, Genoscope, Institut François Jacob, CEA, CNRS, Univ Evry, Université Paris-Saclay, 91057 Evry, France
6Globe Institute, University of Copenhagen, Copenhagen, Denmark
7Mississippi State University, Department of Biological Sciences, Mississippi State, Starkville, 39762, USA
8Natural Sciences Museum of Barcelona, P◦ Picasso s/n, Parc Ciutadella, 08003 Barcelona, Spain
9Department of Biomedical Informatics and Data Science, Yale School of Medicine, New Haven, CT, USA
10Arizona Genomics Institute, The University of Arizona, Tucson AZ 85721, USA
11Wise Ancestors, 241 Kings Village Rd. #66671, Scotts Valley, California 95067, USA
12Vertebrate Genome Laboratory, The Rockefeller University, New York, NY 10065, USA
13Genotek Ltd., Moscow, 105120, Russia
14Centro de Oncologia Molecular, Hospital Sirio-Libanes, Sao Paulo, Brazil
15National Human Genome Research Institute, National Institute of Health, Bethesda, MD, 20892, USA
16Royal Botanic Gardens, Kew, Richmond, Surrey TW9 3AB, UK
17Julie Ann Wrigley Global Futures Laboratory, Walton Center for Planetary Health, Arizona State University, Phoenix, AZ 85281, USA

18Department of Data Science, Dana-Farber Cancer Institute; Department of Biomedical Informatics, Harvard Medical School, Boston, USA
19Department of Medical Biochemistry and Microbiology, Uppsala University, Uppsala, Sweden
20Vertebrate Genomics, Broad Institute, Cambridge, MA,USA
21Department of Molecular Genetics, The Ohio State University, Columbus, USA, 43210
22Departament de Biologia, Universitat de Girona, Girona, Spain
23Departament de Biologia Cellular, Fisiologia i Immunologia, Universitat Autònoma de Barcelona, Cerdanyola del Vallès, Spain
24SequAna Core Facility, Department of Biology, University of Konstanz, Konstanz, Germany
25Department of Biology, University of Kentucky, Lexington, KY, USA
26Department of Biology, University of Florence, Sesto Fiorentino 50019, Italy
27Faculty of Biosciences and Aquaculture, Nord University, Norway
28Vale Institute of Technology, Belém PA 66055-900, Brazil
29Environmental Bioinformatics Group, SIB Swiss Institute of Bioinformatics, Batiment Amphipole, UNIL-Sorge, 1015 Lausanne, Switzerland
30State Key Laboratory of Genome and Multi-omics Technologies, BGI Research, Shenzhen 518083, China
31Center of Evolutionary & Organismal Biology, Zhejiang University School of Medicine, Hangzhou, 310058, China
32Department of Genetics, University of Cambridge, Cambridge CB2 3EH, UK
33State Key Laboratory of Genome and Multi-omics Technologies, BGI Research, Shenzhen 518083, China

*Kerstin Howe kj2@sanger.ac.uk

# Abstract

Reference genome assemblies are foundational resources across the biological sciences as they begin to expose the fundamental building blocks of each species and the molecular toolkits vital for adaptation and survival. High-quality genomes enable insights into evolution, facilitate ecosystem monitoring and protection and conservation of species, access to new biomaterials and biomedicine, advances in agri- and aquaculture, and support planetary health. The amount of information gathered from a species' genome assembly is directly dependent on its quality. However, different sequencing technologies and software solutions generate a wide range of quality outcomes. Here, the Earth BioGenome Project's Sequencing and Assembly committee, together with a large community of researchers performing eukaryote sequencing and assembly, outlines quality standards for a "reference" assembly and formulates recommendations to ensure these standards are met.

# Background

Since the sequencing of the first genomes of fungi, animals, plants, and protists more than 25 years ago, biodiversity genomics has grown to be an active and fruitful field. The Earth BioGenome Project (EBP) is a global network of projects, promoting the generation of high-quality reference genome assemblies for all extant eukaryotic life (Lewin et al. 2018). The EBP brings together regional, ecosystem- and taxon-focussed projects, united in their vision to advance our understanding of the natural world and support planetary health.

The EBP is now scaling up from Phase I, sequencing representatives for the approximately 10,000 species representing half of the eukaryotic families, to Phase II, which will sequence and assemble 150,000 species covering 50% of the 140,000 known genera (Blaxter et al. 2025), including all endangered species, and species of economic and ecological importance. Given the real costs of specimen collection, especially of rare or endangered species, it is critically important to guarantee that each genome assembly produced is the best possible. While sequencing and assembly methods are under continuous improvement, delivering the best possible assemblies with current technologies will future-proof ongoing efforts.

The EBP Sequencing and Assembly Committee was formed to define assembly quality standards and to recommend workflows to achieve these standards, based on the current consensus within the eukaryote genome assembly community (Durbin, Howe, Jarvis, Myers, Phillippy, Rhie, Zhang, et al. 2026; Durbin, Howe, Jarvis, Myers, Phillippy, Rhie, Guojie, et al. 2026). Here we summarise the present best practice for sequencing and assembly to ensure high completeness, contiguity and quality in reference genomes. These findings are embedded within an ecosystem of standards and recommendations for the end-to-end workflow of biodiversity genomics projects generated by the EBP and promote the FAIR (Wilkinson et al. 2016) reporting and publication of all data for fully interoperable use (Lawniczak et al. 2025; Santos et al. n.d.; Howard et al. 2025; Lawniczak et al. 2022; Howe et al. 2021)). We intentionally do not provide explicit requirements for primary data or software to be deployed, as there are competitive alternative technologies available at all steps. Furthermore, technologies develop, both for sequencing and computational methods, and different combinations of data can be used to achieve these goals. However, it is useful to provide concrete recommendations based on the gathered experience, to support emerging new projects and facilities that will be required to meet the EBP Phase II goals.

We strongly encourage all genome projects to adopt the recommendations and standards presented here. However, we recognise that this may not be feasible in every circumstance. Accordingly, the recommendations that follow should be interpreted with the implicit qualification of "where possible." Importantly, failure to meet all recommended standards should not preclude the publication of genomic data, as datasets of varying levels of completeness and quality still provide substantial value to the research community. However, there should be a continuing goal to revisit lower-quality assemblies and bring them up to the target standard, as that becomes practical.

# Assembly process

## Generating sequencing data to support optimal assembly quality

Modern high-quality reference genome sequencing has been built upon innovations in sequencing technologies, in particular the advent of long reads, many kilobases in length, and the development of scaffolding methods based on chromatin conformation capture (Hi-C, Omni-C, Pore-C, CiFi and related methodologies). Achieving the proposed genome assembly quality standards requires both data types (Durbin, Howe, Jarvis, Myers, Phillippy, Rhie, Zhang, et al. 2026). **Table 1** summarizes current state-of-the-art sequencing technologies and recommended coverage for genome assembly. DNA extraction methods suitable for long read sequencing are under active development, and a compendium of best practice across a wide span of phylogenetic diversity has recently been published (Howard et al. 2025).

We recommend that each species' genome assembly be generated from a single individual. This reduces the assembly to the haplotypes present in the individual chosen. For species that reproduce clonally, such as cultured protists, vegetatively propagated plants, fungi and many parthenogens, multiple individuals from a clonal colony or clonally derived culture can be used. For cultured, sexually-reproducing organisms where the mass of DNA available in a single individual is not enough for the input requirements of the sequencing technologies, inbreeding through isolated brother-sister mating for several generations can reduce the haplotypic diversity present and improve assembly performance. The inbred culture should be retained (e.g. by deposition in a culture collection) to assure availability for post-genomic work.

When available material is extremely limited, for example from minimally invasive biopsies or small-bodied organisms, DNA amplification can enable long-read sequencing from otherwise insufficient DNA quantities (Laumer 2022). DNA amplification can also help if co-extracted substances, such as mucus or naturally occurring metabolites, inhibit long read sequencing (Bein et al. 2025; Howard et al. 2025).

In many species, sex is determined through the presence of differentiated sex chromosomes, and we strongly recommend sequencing the heterogametic sex (males in XY, females in ZW systems). In UV systems, where sex is determined in the haploid phase and there is no heterogametic sex, the diploid sporophyte carries both the U and V chromosomes and should be preferred for sequencing where this stage is accessible (Coelho et al. 2018). Where the presence of supernumerary chromosomes (B chromosomes) is suspected, efforts should prioritize sampling an individual known or suspected to carry these genetic elements. Where sex determination is haplo-diploid, or the species have large, persistent haploid phases, the haploid sex or stage can be sequenced to make full assembly easier. Where phenotypic morphs are associated with supergenes (large, non-recombining chromosomal variants), sequencing should target a heterozygous individual that carries both chromosomal variants (examples in (Jeong et al. 2022; Tuttle et al. 2016). In the majority of species the somatic and germline genomes are identical and whole organisms can be processed for sequencing, but some species have germline-restricted DNA or chromosomes, for example the germline chromosomes of passerine birds and fungus gnats, or the programmed DNA elimination observed in ascaridid nematodes (Sotelo-Muñoz et al. 2022; Stevens et al. 2026; Hodson et al. 2022). In these cases, care should be taken to sequence tissues where the germline-eliminated DNA is present. In ciliate protozoa, each dikaryotic cell contains a germline nucleus (the micronucleus) and a somatic nucleus (the macronucleus, derived from deletion and rearrangement of the germline genome). Nuclear sorting from cultures can be used to generate a germline-only preparation. Genome rearrangement is also used in control of gene expression in some somatic cell lineages (such as the B-cell and T-cell receptor gene families of mammals) and assembly of genomes from somatic tissues where rearrangement is known to occur should be aware of the impact of sequencing the multiple, independently-rearranged versions of these loci.

Technology developers strive to make their sequencing methods universal, able to sequence the complete genome from all species. However, some genomic regions or taxa present platform-specific challenges. Pacbio HiFi data in general yields assemblies with higher sequence accuracy, resulting in a better separation of the haplotypes (Holt et al. 2024; Lang et al. 2020). It excels for assembly of most repeat-rich genomes where minor variation between repeats permits accurate reconstruction of tandem arrays. Oxford Nanopore Technologies (ONT) reads may provide better coverage or contiguity for particular regions, such as highly GC-rich avian and reptile microchromosomes, and can replace or complement HiFi data when near-complete or telomere-to-telomere assemblies are required (Smeds et al. 2025; Li and Durbin 2023). ONT also appears particularly advantageous for repeat-rich teleost fish genomes and several invertebrate groups, including gastropods and crustaceans, whose genomes can contain abundant low-complexity regions and simple sequence repeats, such as long dinucleotide repeats. The abundance of these genomic features can be estimated from independent genomic data and used to guide technology selection before large-scale sequencing (https://github.com/Genoscope-LBGB/freq-motif-fastq). The technology manufacturers have recognised these issues and strive to rectify them.

Ideally long range scaffolding data should be generated from the same individual used for long read data generation. If that is not possible it should be derived from another individual or even a pool of additional individuals of the same species, preferably of the same sex and from the same population as the focal organism.

Short read sequencing of parents can be used to fully phase the genome sequence of offspring (Koren et al. 2018). Ideally both maternal and paternal genomes should be sequenced. In cases where only one parent can be identified with certainty, generating short or long read sequencing of that one parent can assist with accurate phasing.

We further recommend the generation of at least one transcriptome dataset for each species, to aid in gene structure prediction during annotation of the genome. These data can be short read (e.g. Illumina-based RNAseq) or long read (PacBio IsoSEQ or Kinnex, ONT long cDNA or direct RNA seq). Multiple transcriptome datasets from different tissues are of increased utility (https://www.earthbiogenome.org/report-on-annotation-standards).

Generating data from additional technologies can improve genome assembly during the scaffolding stage. Physical and optical restriction fragment maps can be used to refine the order and orientation of assembled scaffolds. Genetic maps, from reduced representation sequencing or other sequence-based marker sets can order, and orient if marker density is high enough, contigs along chromosomes. Genetic maps can be particularly useful for correct assembly of species with haplotypes that differ by major structural rearrangements. Karyotyping and fluorescent *in situ* hybridisation (FISH) can be used to confirm chromosomal numbers and assignments and the structures and positioning of centromeres and other structures. These technologies are useful additions to genome assembly workflows, but are not essential.

**Table 1**: Current state-of-the-art sequencing technologies for genome assembly.

| ***Name*** | ***Providers*** | ***Characteristics*** | ***Requirements for reference assembly*** |
|---|---|---|---|
| **Long Read Data** | | | |
| HiFi | Pacific Biosciences | ● very high accuracy (Q30-40)<br>● typically 10-20 kb length | ● >15x coverage per haplotype |
| ONT | Oxford Nanopore Technologies | ● moderate to high accuracy using the R10.4 or later chemistry (Q20-30)<br>● length from 10 kb to 1,000 kb with yield decreasing with higher mean length<br>● can be assembled into comparable quality outcomes without the need for additional separate error correction | ● >25x coverage per haplotype of reads with N50 ~50kb<br>● error correction of this data using herro or hifiasm --ont<br>● alternatively add >25x coverage per haplotype of short read for polishing |
| CycloneSEQ | MGI | ● moderate to high accuracy using S1 chemistry (Q20+)<br>● standard read N50 20–50 kb;<br>● Ultra-long scheme achieves N50 >100 kb | ● >25x coverage per haplotype of reads with N50 ~50kb<br>● error correction of this data using herro or hifiasm -ont<br>● alternatively add >25x coverage per haplotype of short read for polishing |
| **Long Range Data** | | | |
| Omni-C | Dovetail Genomics | ● DNase / sequence-independent endonuclease digestion, | ● >25x coverage per haplotype |
| | | reducing bias and even coverage across the genome | |
| Hi-C | Arima Genomics | ● clade specific multi restriction enzyme based Hi-C | ● >25x coverage per haplotype |
| Hi-C | Phase Genomics | ● clade specific multi restriction enzyme based Hi-C | ● >25x coverage per haplotype |
| Hi-C | Quiagen | ● clade specific multi restriction enzyme based Hi-C | ● >25x coverage per haplotype |
| Pore-C | Oxford Nanopore Technologies | ● protocol (not commercial kit)<br>● long-read sequencing of restriction-enzyme based chromatin conformation capture concatemers<br>● better performance in repetitive or structurally complex regions | ● >5-10x coverage per haplotype |
| CiFi | Pacific Biosciences | ● protocol<br>● long-read sequencing of restriction-enzyme based chromatin conformation capture concatemers<br>● better performance in repetitive or structurally complex regions<br>● low input requirements (60,000 cells) | ● >5-10x coverage per haplotype |
| Ultra Long ONT | Oxford Nanopore Technologies | ● 100 kb or longer read lengths | ● 20x coverage per haplotype |

## Recommended assembly strategies

Below we give an outline of recommended assembly strategies, based on community experience, with options for the different data types. Due to the number of software tools we mention, their references are listed in **Suppl. Table 1** rather than in the main document.

An overview of the assembly generation process is given in **Table 2.**

**Table 2:** Computational workflow steps for assembly generation. The publicly available outputs are coloured blue. Detailed description in the text.

| ***workflow step*** | ***action*** | ***output*** |
|---|---|---|
| 1. sample registration | permit acquisition | |
| | taxonomy ID generation/retrieval | |
| | ToLID registration | ToLID |
| | BioSample registration | Biosample accession |
| | BioProject registration | BioProject accession |
| 2. genome characteristics | coverage calculation | |
| | genome size assessment | |
| | heterozygosity assessment | |
| | ploidy identification | |
| | karyotype identification | |
| 3. contig assembly | assembly | |
| | duplication removal | |
| | organelle assembly | organelle assemblies |
| 4. scaffolding | scaffolding | draft assembly |
| 5. cobionts | separation of target species from cobionts and contamination | |
| | assembly of cobionts were possible | cobiont assemblies |
| 6. polishing | usually not necessary | |
| 7. curation to chromosome level | removal of remaining non-target species sequence | |
| | removal of remaining erroneous duplications | |
| | correction of order and orientation | |
| | identification and naming of chromosomal | final reference assembly |
| | units | |
| 8. assembly metrics | calculation of genome metrics | |
| 9. INSDC submission | generation of data BioProjects<br>submission of raw reads<br>submission of primary haplotype/haplotype1 including organelle assemblies<br>submission of additional haplotypes<br>submission of any cobiont assembllies | full set of submitted project data |

## 1. Data system organisation and registration

Careful design of the data structures used to hold the raw sequencing and assembly products are essential in both management of the sequencing project and in communicating the results on publication. The EBP uses the National Center for Biotechnology Information (NCBI) Taxonomy Database (TaxonomyDB) for species names (Schoch et al. 2020). This is not because TaxonomyDB is more definitive than other synoptic taxonomic systems, but rather because it is the one most closely related to the public release of genomic data. If your species does not have an entry in TaxonomyDB, it is simple to register it (see (Blaxter et al. 2024)).

After obtaining one or more BioSample IDs for your sample, they should be associated with a species-level umbrella BioProject, and child data BioProjects generated for each raw data type (Figure 2). The assembly should be submitted under its own child data BioProject under the species-level BioProject. If the data belongs to an EBP affiliated project it should be linked to the EBP umbrella BioProject (accession PRJNA533106).

To ease communication of assembly results, the EBP has adopted the ToLID (Tree of Life identifier) system, which offers a unique, easily-communicated, short (and human-readable) “name” for each specimen. Specimens selected for sequencing should be registered for a ToLID at [id.tol.sanger.ac.uk](id.tol.sanger.ac.uk) and assemblies named after this ToLID (see “INSDC submission: project structure and nomenclature” section below for further information).

## 2. Genome characteristics

Prior assembling any whole genome sequencing data, an analysis of the k-mer spectra of the raw read sets should be performed (Jenike et al. 2025). These approaches use raw sequencing reads and estimate sequencing coverage, genome size, heterozygosity and ploidy, critical variables to consider during assembly. There are multiple tool-chains for this, e.g. **KMC**; **FastK** followed by **GenomeScope 2.0**; with **Smudgeplot** for determining ploidy.

Issues can arise in autopolyploid species that share considerable sequence similarity between the subgenomes present, leading to underestimates or incorrect predictions of genome size and ploidy level. Substantial presence of non-target species in the sequenced sample can also complicate the interpretation.

The Genomes on a Tree (GoaT) datasystem (Challis et al. 2023)) collates genome size estimates generated using non-sequencing methods (flow cytometry or densitometry) as well as the assembly sizes of genomes submitted to the International Nucleotide Sequence Database Collaboration (INSDC: NCBI GenBank, the European Nucleotide Archive and the DNA Database of Japan (Karsch-Mizrachi et al. 2025)). Where values are not available for a species, GoaT estimates genome size on the basis of actual values from the closest relatives with determined values. These values can be used as a ground truth (directly measured or assembled values) or a guide (estimated values) against which to compare the values derived from k-mer distributions. A collection of karyotype sources can be found in **Suppl. Mat Table 2**.

## 3. Contig Assembly

During contig assembly long reads are used to derive stretches of contiguous sequence without gaps or ambiguities. Modern assemblers first generate a sequence graph that represents all the overlaps between the input raw data before traversing the graph, splitting the graph at forks and emitting contigs. Where there is heterozygosity and the level of sequence divergence (allelic polymorphism) allows, they emit separate contigs for the two haplotypes (in a diploid; multiple haplotypes in polyploids) rather than a collapsed representation of both alleles. We strongly recommend the use of **hifiasm** or **Verkko2**. If parental read data are available, assemblers can be used in trio-binning mode, and the emitted haplotypes are grouped by likely maternal or paternal origin (mat/pat). The assemblers assess read coverage of the contigs, and use this and sequence similarity to attempt to separate primary and alternative haplotypes into two bins (pri/alt [primary, alternative] or hap1/hap2). Hi-C data can be involved at the contig stage to provide phasing, and assemblers emit a number of (pseudo)haplotypes according to a preset ploidy. Even when the assemblers do this, it may be necessary to remove remaining haplotypic duplicates from all generated (pseudo-)haplotypes using **purge_dups**, before scaffolding. K-mer histograms and duplication levels in Benchmarking Universal Single-Copy Orthologs (**BUSCO**) analyses can aid the decision (Tegenfeldt et al. 2025).

Most eukaryotes carry organellar genomes from the mitochondrion (all taxa) and the plastid (photosynthetic taxa and derived groups). These genomes should be assembled independently, using organelle-specific toolkits such as **oatk** which deal with high coverage, heteroplasmy, circularity and frequent overlap with nuclear insertions of organellar DNA. Animal mitochondria, which tend to be less complex, can be assembled with **MitoHiFi** (for HiFi data) or **MitoVGP** (for ONT or PacBio CLR data).

## 4. Scaffolding

Scaffold with long range Hi-C or Omni-C data using **YaHS**. YaHS is recommended over older tools such as **SALSA2**, as these have demonstrated limitations (Zhou et al. 2023). For Omni-C data, **HiRise** or **3D-DNA** can also be used. If using long-read CiFi or Pore-C data, these can be aligned with **wf-pore-c** or **Falign** before using YaHS. Both hifiasm and Verkko2 are able to incorporate long-range (Hi-C, Pore-C or CiFi) data during contig assembly, and thus can emit phased haplotypes.

## 5. Cobionts

Samples taken from the wild may include cobionts: organisms that are not the primary target of the genomic sequencing project, such as microbes in the gut microbiome, mutualist symbionts, pathogens, parasites and chance contaminants. It is important to leave only sequences from the target species in the final assembly, and positively identify and remove sequences belonging to the other cobionts present in the sample. Tools such as **BlobToolkit, FCS-GX** , **ASCC** and **Kraken2** can help identify and mark these for removal. We recommend applying them after the scaffolding stage. Where abundant cobionts are detected, it may be advisable to repeat the contig assembly and scaffolding after removal of identified cobiont genomes, to assure best performance.

In addition to the target species, we recommend separately assembling and releasing the identified cobiont genomes. The process should follow the above, e.g. using **hifiasm**, for eukaryotic cobionts, or a metagenome assembler such as **metaMDBG** or **Myloasm** to assemble microbial communities. Quality checks can be carried out using **BUSCO** or **CheckM2**, and for prokaryotic Metagenome-Assembled Genomes (MAGs) the assemblies can be validated against the MiMAG standard (Bowers et al. 2017). The additional data should be submitted alongside the target species' assembly as detailed in Figure 2.

## 6. Polishing

If the assembly was constructed from HiFi data, or corrected ONT data (prior to assembly or using hifiasm –ont) polishing is not necessary. If the contigs were constructed from pre-2024 Oxford Nanopore chemistries (e.g. pre-R10), or Continouse Long-Read (CLR) data from PacBio chemistries, polishing of the assembly may be required depending on the quality value (QV, see below). We recommend using short read data from a high accuracy platform (such as Illumina or Element) and **bwa** (to map), **FreeBayes** or **DeepVariant** (to variant call), **Merfin** (to filter calls) and **bcftools** (to apply corrections). If further polishing is required/desired, this can be done by remapping the long reads with **WinnowMap**, calling variants with **DeepVariant**, and then correcting as above. Correction with short reads may improve accuracy, particularly for insertion-deletion (indel) errors, but care must be taken to avoid inserting errors due to uncertainty of short read mapping and simply “correcting” the long read data to represent the error biases of the short read platform. It is important to map the polishing reads to all of the assembly material (primary, alternate, contamination, organelles) to avoid mismapping and false calls.

## 7. Curation to chromosomal level

The process above will generate assemblies, but our experience has shown that these assemblies still contain errors, which are relatively easy to identify and correct before submission (Howe et al. 2021). Curation turns a scaffolded assembly into one where the majority of the assembled sequence is ordered and oriented in chromosome-scale pseudomolecules. While artificial intelligence approaches have been reported to speed up the process (Jiang et al. 2024; Mergez et al. 2026), curation is still largely manual, relying on visual inspection and expert assessment of the Hi-C contact map.

For curation, the Hi-C data is re-mapped to the assembly (including all haplotypes where these have been assembled), and the order and orientation of scaffolds is reviewed in **PretextView** or **JuiceBox**. Workflows supporting curation are available (GRIT Rapid Curation, TreeVal, Galaxy, see below). The curation of micro- and dot chromosomes as present in e.g. bird and reptile genomes benefits from the additional use of tools like **MicroFinder.**

While most sequences can be assigned to chromosomal pseudomolecules (super-scaffolds), some small contigs or scaffolds will remain unplaced or are impossible to localize inside the super-scaffolds. As they are part of the genome, these sequences should be submitted as unplaced or unlocalised according to INSDC terminology, however the goal is for these to represent less than 10% of the assembly sequence. The curation process can correct errors in order and orientation of scaffolds and phasing errors, remove redundant haplotypic contigs and assign chromosome numbers and names to chromosomal scaffolds. The latter should follow established data/nomenclature where it exists for the target species or simply number the chromosomes by size. More detailed descriptions of the curation process can be found elsewhere (Howe et al. 2021; Wood et al. n.d.)

As described above, for species that have heterogametic karyotypes (X0, XY, ZW, and variants thereof) sequencing of the heterogametic sex (male or female as relevant) is strongly recommended. Because it is desired that the primary reference contains as complete a representation as possible of the species' genome in haploid form, we recommend that .hap1 should contain the entirety of all known sex chromosomes, along with metadata describing the location of any pseudoautosomal regions and/or the non-recombining region to allow for downstream masking of the recombining sequences, and prevent loss of information. An important requirement is that all the material should come from the same individual, and that the sum of the material in all haplotype assemblies should give the complete diploid or polyploid genome. For X0 or Z0 species, .hap1 should contain the X or Z chromosome. For U/V species, .hap1 should include the U or V; however sequencing one sexed individual (male or female) will likely preclude the confident identification of the sex chromosomes.

We suggest adding annotations of putative sex chromosomes to the assembly description, and we are working with INSDC to possibly implement a nomenclature indicating the contingent nature of these identifications that can be directly applied as chromosome name.

In some cases, where an individual of the homogametic sex has been sequenced, a heterogametic sex chromosome assembly derived from another individual has been added to a submission to make a reference for mapping, e.g. adding a Y from a male when the sequenced individual was an XX female. We advise against this and propose instead that a synthetic reference for alignment is made at point of need using material from two different primary assembly submissions. This also applies to UV species, where a synthetic reference should contain one haplotype and both the U and the V chromosomes.

## Diploid and polyploid assemblies

It is now possible to phase or separate out homologous chromosomes during assembly to provide two (or more, in polyploids) essentially complete sets of chromosomes. There are two standard approaches to this. The first uses sequence data from the parents to fully phase the chromosomes into diploid sets, resulting in a maternal haploid genome and a paternal haploid genome, and a fully parentally phased assembly. We recommend that these are called <tolid>.mat and <tolid>.pat and that an additional synthetic genome including both X/Z and Y/W for annotation purposes is provided. The second uses long range phasing data from Hi-C or (ultra)long reads to phase individual chromosome assemblies. This potentially generates fully phased individual chromosomes, or ones with few phase switches, but no phase coherence across chromosomes. In the latter case we recommend that the resulting haploid genome sets are called <tolid>.hap1 and <tolid>.hap2. For polyploids, e.g. tetraploids, the corresponding nomenclature would be <tolid>.mat1, .mat2, .pat1, .pat2 or .hap1, .hap2, .hap3, .hap4, etc. We suggest that reasonable efforts are made where possible to assort the relevant subgenomes of polypoloids into submitted haplotypes. This can e.g. be based on kmer similarity between repeats and transposable elements.

Despite submitting two or more full haploid genomes that are effectively equivalent in quality, the wider community still for many purposes uses a single “primary” haploid reference genome sequence. By default this should be .hap1 if phased with long range data. For trio-phased assemblies the primary assembly will need to be designated (or a synthetic reference constructed that contains both heterogametic sex chromosomes).

The organellar (mitochondrion, plastid etc.) genomes should be included in the primary or .hap1 assembly.

## Telomere to Telomere (T2T) assembly

The recommendations above are designed to achieve the EBP assembly quality target metrics as defined below. Recent developments have made yet another approach possible: the generation of complete **telomere-to-telomere** (T2T) assemblies (Li and Durbin 2023; Nurk et al. 2022; Hansen et al. 2026; Formenti et al. 2026). These are usually based on high quality HiFi and ONT long reads combined with ONT ultralong reads, require significant resources and are therefore not suggested to replace the EBP reference quality standard at this point in time.

Current recommendations are to generate >30 fold coverage in PacBio HiFi sequence and ideally >20 fold coverage in ultralong (>100kb) ONT sequence, in addition to deep (>25 fold coverage) Hi-C for confirmation, for each haplotype. A dedicated assembler for this configuration is **Verkko**, which has also been proven successful in non-human species. Recent versions of **hifiasm** also support integration of HiFi or regular ONT, ultralong ONT and Hi-C data. There are also investigations into an ONT-only recipe including PoreC, using the **SQK-APK114, SQK-LSK114, and SQK-ULK114 pipeline**. At present, for large genomes, assembling a truly T2T reference genome usually requires the use of an assembler that preserves the graph structure, which enables manual graph resolution of the complex regions using **TTT** and/or **Verkko-fillet**. This has been shown convincingly in Verkko.

## Workflows

Workflows developed using the approaches and tools listed above are available in WorkFlowHub (Gustafsson et al. 2025), including collections from EBP affiliates such as the Vertebrate Genomes Project (VGP; (Rhie et al. 2021)), the European Reference Genome Atlas (ERGA; (Mazzoni et al. 2023)), the Sanger Tree of Life programme (ToL; https://www.sanger.ac.uk/programme/tree-of-life/ ), and the ATLASea marine genomics project (https://www.atlasea.fr/en/).

- VGP: https://workflowhub.eu/collections/8
- ERGA: https://workflowhub.eu/programmes/33
- Sanger Tree of Life: https://workflowhub.eu/programmes/37
- ATLASea: https://workflowhub.eu/collections/41

Other pipelines are waiting to be added (e.g. Genomics of Brazilian Biodiversity: https://github.com/itvgenomics/pipeasm)

# Assembly standards

Compliance with the EBP assembly standards requires assembly generation and curation according to the processes described above. Below we detail the measurement and assessment of the resulting assembly.

## Quantitative assembly standards

Quantitative assembly metrics can easily be reported by using the following nomenclature: X.Y.Qn with X denoting the $\log_{10}$ of the contig N50 (rounded down and reported as a single digit), Y denoting the $\log_{10}$ of the scaffold N50 and n denoting the base call quality value. If chromosomal scale N50 is achieved, X and/or Y can be reported as “C”. For instance, 6.C.Q40, describes a megabase N50 contig contiguity and chromosomal scale N50 scaffolding, with less than 1/10,000 error rate.

We have summarised the standard metrics for different organism groups in Table 3 and tools for measuring the suggested assembly metrics in Table 4. An approximation of the distribution of EBP metrics over currently available EBP and non-EBP genome assemblies can be found in Figure 1.

**Table 3**: Quantitative assembly standards. Note the caveats around completeness and single copy genes identification in the text.

| *Species group* | *DNA availability* | *Assembly standard* | *Additional requirements* |
|---|---|---|---|
| **Standard material availability** | >0.5µg | 6.C.Q40 | >90% assigned to candidate chromosomal sequences<br>>90% kmer completeness<br><5% false duplications<br>>90% single copy genes<br>>90% transcripts mappable |
| **Standard material availability - T2T** | >40µg | C.C.Q60 | |
| **Limited material (ultra low input)** | <0.1µg | 5.C.Q40 | |
| **Extremely limited material** | <0.001µg | 5.6.Q40 | >90% kmer completeness<br><5% false duplications<br>>90% single copy genes<br>>90% transcripts mappable |
| **Unculturable single cell eukaryotes** | variable | none as yet | none as yet |

**Table 4:** Tools for measuring the suggested assembly metrics. Note that Merqury requires a high accuracy shotgun data set (Illumina, ONT data corrected with herro (https://github.com/lbcb-sci/herro) or PacBio HiFi data) to extract kmers from. Ideally this will be independent from the one used to build the primary assembly. Validation of haplotype phasing accuracy requires genome-wide sequence from a close relative so is not possible in many cases. It should be carried out where possible, but is not required to meet the EBP assembly standard.

| *Quality parameter* | *Metric* | *Methods* |
|---|---|---|
| Continuity | Contig (NG50) | Genometools seqstat or equivalent<br>NG50 requires a genome size estimate, obtainable from GenomeScope, PreQC, or other tools |
| | Scaffolds (NG50) | |
| | Gaps / Gbp | |
| Structural accuracy | False duplications | Merqury |
| | Reliable blocks | Raw data mapping + Asset + script to accumulate desired blocks |
| | Curation improvements | False duplication removal: purge_dups |
| Base accuracy | Base pair QV | Merqury, MERQURY.FK, yak |
| | k-mer completeness | Merqury, MERQURY.FK, yak |
| Haplotype phasing | Phased block (NG50) | Merqury, with availability of parental / siblings / population illumina data |
| Functional completeness | Genes | BUSCO, OMArk, compleasm |
| | Transcript mappability | RNA-seq preprocessing + STAR or other dedicated transcript mappers |
| | Transposable elements | LTR_retriever (LAI) |
| Chromosome status | Assigned % | Assigned during curation |
| | Sex chromosomes | Assigned during curation |
| | Organelles (e.g. MT) | Assigned during curation |

## Eukaryotic species with sufficient tissue and DNA available.

The standard long read and long-range (Hi-C) methods require input DNA masses of 0.5 to 10 µg. For most larger species this yield is readily achievable from a single individual. For these species we propose a minimum reference standard of **6.C.Q40**. Where a telomere-to-telomere assembly has been achieved, with no gaps, the contig value can also be cited as "C".

Alongside these contiguity and error rate goals we propose the following additional criteria

- > 90% sequence should be assigned to candidate chromosomal sequences
- > 90% kmer completeness (i.e. >90% of the kmers present in the reads should be in the assembly)
- < 5% false duplications
- > 90% single copy conserved genes (e.g. the relevant taxonomic BUSCO set) complete and single copy
- > 90% transcripts (or transcriptome reads) from the same organism mappable

Links to standard tools for measuring these metrics are provided in Table 3.

As with all assembly quality metrics, these thresholds should be interpreted in the biological context of the target species. For example, k-mer completeness may be affected by contamination, symbionts or other non-host DNA, particularly in taxa such as sponges and algae. Consequently, values below the recommended threshold do not necessarily indicate an incomplete assembly and should be interpreted alongside contamination estimates and complementary quality metrics. Similarly, BUSCO scores may be unrepresentative in situations where the target organism diverges from the set of organisms used to build the BUSCO lineage dataset, or in case of incompletely rediploidized polyploid organisms typically showing high levels of BUSCO duplication (Alam et al. 2025). This will also be evidenced in species with large genomes exhibiting high numbers of repeats containing paralogs, as well as species with active retrotransposition (Heckenhauer et al. 2022).

T2T assemblies are included in our definition of standards. However, since the moniker "T2T" is often generously used in the literature (see Rhie et al., 2021 for a definition), we suggest that it should be applied only when all telomere sequences are present, i.e. there is at least the start of the true telomeric repeat on each end of a linear chromosome, there is a complete representation of the sequence between each chromosome end, i.e. there are no sequence gaps within chromosomes, and ideally the base-level QV is greater than 60 (i.e. the assembly will be scored **C.C.Q>60**).

### Species with limited DNA or material per individual.

For many species the maximum theoretical yield from one individual is less than 100 ng DNA, and this currently precludes processing through standard library construction and sequencing methodologies. New whole genome amplification technologies that use highly processing DNA polymerases (Ampli-Fi) allow data generation sufficient for assembly from as little as 1 ng of DNA (Bein et al. 2025). Hi-C generation from a second individual or pooled individuals can be used for scaffolding. For such species (where total DNA mass is below 100 ng per individual), we propose that a standard of **5.C.Q40** is regularly achievable, i.e. similar to the larger sample target but with contig N50 relaxed to >100kb to reflect amplification dropout.

Advances in sequencing of individual meiofauna (where total DNA mass is <1 ng), following protocols such as PiMmS and phibeta29 replicase whole genome amplification (WGA), now allow generation of assemblies with a 100 kb contig N50 (Laumer 2022; Silander and Saarela 2008). These assemblies can then be scaffolded, sometimes to chromosomal contiguity, with Hi-C data derived through low-input methods from a pool of individuals. We suggest a minimum of **5.6.Q40** for such species, but encourage achieving of **5.C.Q40** where possible.

### Unculturable single cell eukaryotes.

Single cell eukaryotes pose difficulties in both available sample material and genomics architecture. Therefore, no standards have been formulated yet, although the EBP Sequencing and Assembly Subcommittee is currently working toward that goal.

## Additional quality requirements

As already described above, automated processes currently retain multiple errors in the resulting assemblies. We therefore propose that a set of quality control criteria should be reported including:

- separation of sequence of the target species from contaminants and other organisms such as symbionts, members of the gut microbiome, pathogens and parasites (collectively “cobionts”)
- explicit identification of a primary (haploid or pseudo-haploid) assembly, with additional sequence in a secondary bin that may contain either full alternate haplotypes or a set of haplotypic/other sequences from the individual. See below for further discussion of submission structures
- separation and explicit identification of organellar genomes
- only A,C,G,T and N bases present and sequences should not begin or end with Ns

We also encourage:

- identification of discordances between raw data and resulting assembly to locate and remove structural errors (misjoins, missed joins and false duplications)
- identification and naming of chromosomes, esp. sex chromosomes, where possible (see below)

- reconciliation with the known karyotype where it exists and this is possible
- identification of [within individual] PAR (pseudoautosomal region) annotations of sex chromosomes in the form of annotation and/or softmasking of the W/Y PAR, while recognising that this may be difficult in practice for many species
- assembly and submission of all cobionts present at sufficient abundance in the sequenced sample
- > 50% telomere/telomere-like structure completeness, recognising that not all species possess canonical telomeres;
- a read coverage profile that is uniform and coherent with the type of assembly (primary + alts, hap1/hap2, …) submitted
- if possible, orient the chromosome sequence with the p-arm first with consistent orientation between haplotypes.

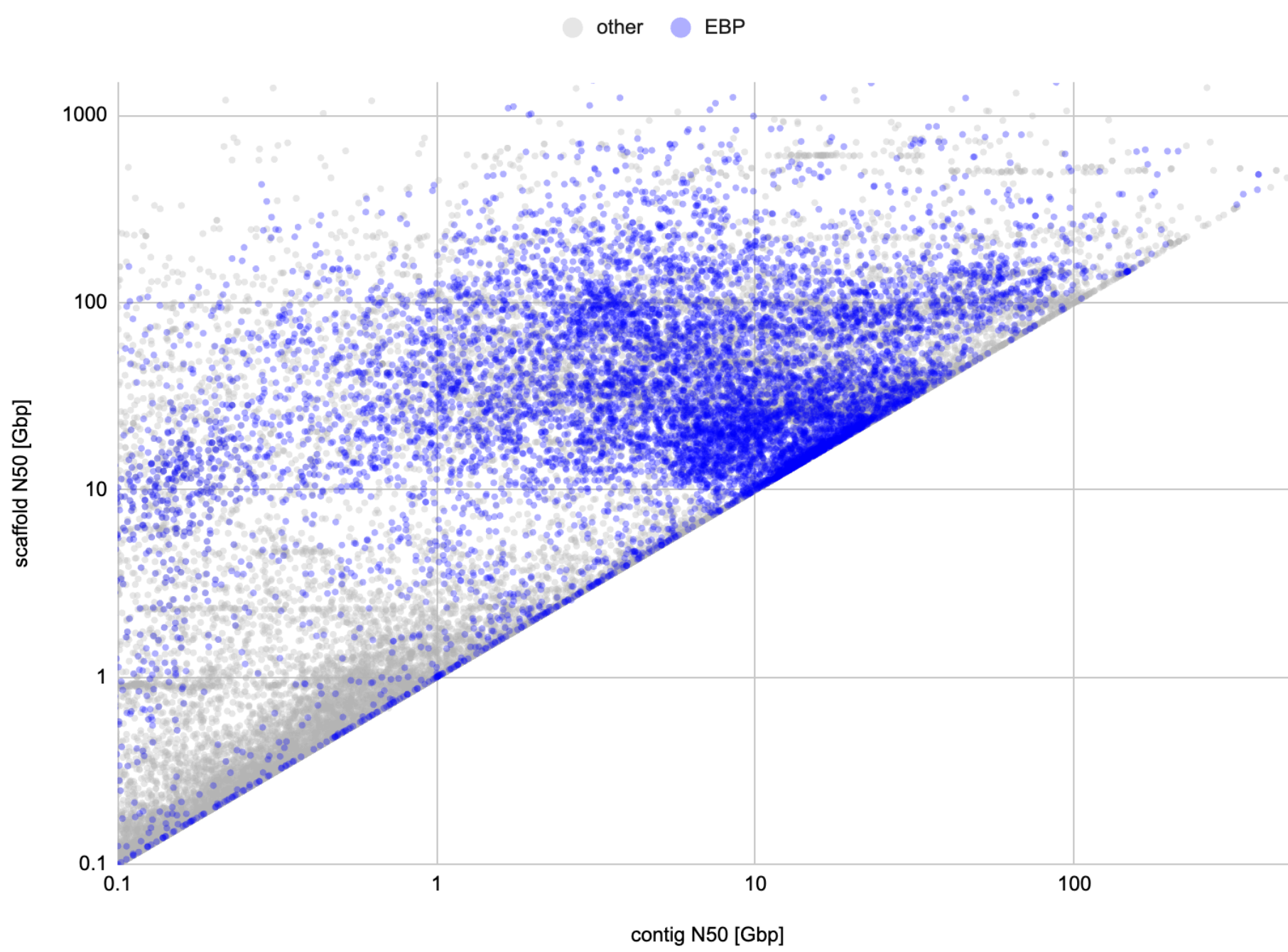

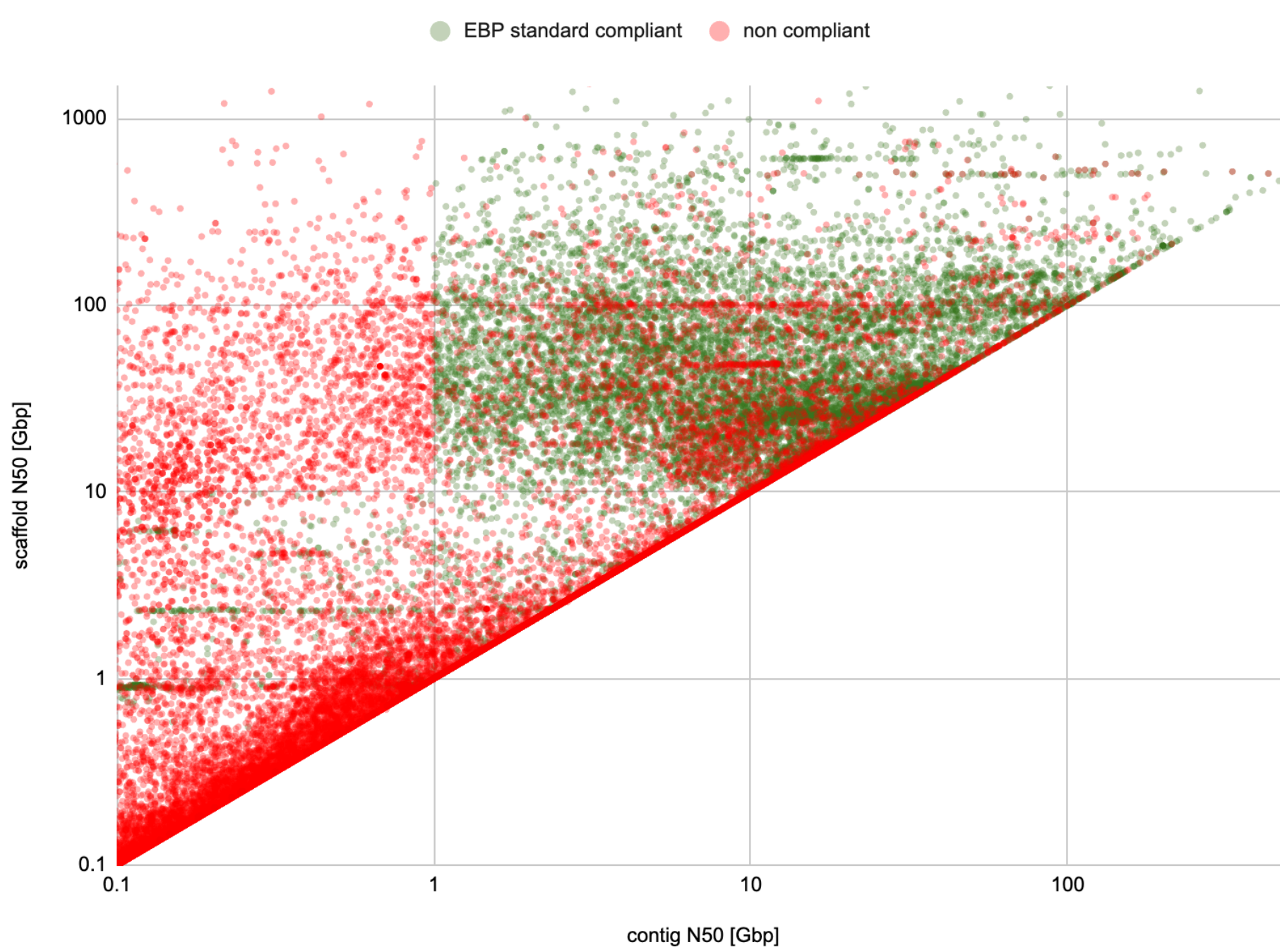

**Figure 1:** **<u>Top</u>**: Contig N50 vs scaffold N50 for EBP and non-EBP assemblies retrieved by querying GoaT for non-human Eukaryota at species rank, excluding assembly_type alternative-pseudohaplotype. It is visible that not all EBP assemblies achieve the targeted

metrics of >1Mb contig N50 and >10Mb scaffold N50, due to small genome size and/or limitations in available material. They were not and should not be excluded from submission. Note that for historical reasons, some short-read based sequencing projects have submitted their projects under the EBP umbrella; these have low contig N50s and don't comply with the standards. **Bottom**: Same assemblies, coloured according to compliance with simplified EBP standards as defined by GoaT (https://github.com/genomehubs/goat-data/wiki/5.1-Earth-Biogenome-Project-(EBP)-FAQs#how-are-ebp-standard-criteria-and-dates-calculated) Assemblies can be compliant outside of the targeted metrics for reasons described above. Also, inside those metrics can be non-compliant if they were e.g. not assembled to chromosome level. Horizontal lines appear where repeated assemblies of the same species resulted in chromosome-sized scaffold N50s at varying contig N50s. This is common for frequently assembled species like crops or model organisms. E.g. the uppermost horizontal accumulation of scaffold N50s around 614Mb belongs to repeated assemblies of *Hordeum vulgare* and the prominent line around 100Mb to *Bos taurus*.

# INSDC submission: project structure and nomenclature

Every genome assembly should be submitted to the INSDC (GenBank/EMBL/DDBJ) Genomes Division for open access use by the scientific community (Karsch-Mizrachi et al. 2025).

When this submission is made the assembly is associated with a "data" BioProject object, which can be part of a hierarchy by being assigned to an "umbrella" BioProject. We suggest the structure in Figure 2 below, with a data project for the raw data and one each per assembly, an umbrella project for the target species (note that under this there may be assemblies of separate symbiont or cobiont species), and then above an umbrella BioProject corresponding to the overall project. EBP projects should link further up to the EBP BioProject object, whose identifier is PRJNA533106 (https://www.ncbi.nlm.nih.gov/bioproject/533106). If the assembly is also contributing to other larger scale efforts such as e.g. GIGA, ERGA or VGP then you can also link your assembly to their umbrella BioProjects.

As well as connecting to a BioProject, an assembly needs to be assigned to a taxonomy ID (txid) entry in the NCBI Taxonomy database (https://www.ncbi.nlm.nih.gov/datasets/taxonomy/, (Blaxter et al. 2024). Although taxid identifiers can be created for informal taxa such as *Maylandia sp. "pearly"*, we would prefer genomes to be associated with taxonomically valid species names, and urge genome projects to work with appropriate taxonomists to identify samples to a species and where necessary to establish the species name in the standard manner in the literature. For EBP assemblies, txid identifiers at levels below species (e.g. subspecies or strains) will be tracked by the EBP at the species level.

Furthermore, in addition to the numerical identifiers generated for assemblies by the public databases we request projects to adopt the ToLID (Tree of Life ID) standard short nomenclature for samples and assemblies, as used by major EBP projects. For individual specimens this takes the form of <clade><gen><spec><ind> where the <ind> is a sequentially assigned integer. All samples from this individual take the resulting ToLID which forms a convenient, human readable identifier used in both data and laboratory settings. For assemblies this takes the form <clade><gen><spec><ind>.<assembly> e.g. ilAlcRepa2.1 for the 1st assembly of **i**nsecta **l**epidoptera *Alcis repandata* individual 2. Unique species designations have been generated to cover ~500,000 species with data in INSDC; other species can be added on request. ToLIDs can also be assigned to subspecies, variants or other taxonomic levels below "species". The assigned ToLIDs will reflect the parent species to preserve the ability of EBP species level accounting. Where a pool of specimens was sequenced, a unique ToLID can be assigned to the pool. A server to view and assign unique individual identifiers for samples is available at https://id.tol.sanger.ac.uk/, where you can also find details on the two letter prefix assignments which assigns all 26 letters to a top level partition of the tree of life.

Finally, it is essential to submit to INSDC the primary raw data used to build the assembly along with the primary assemblies. For ENA submissions, the raw data needs to be referred to in the RUN_REF fields of the manifest of the assembly submission, as indicated in the instructions at https://ena-docs.readthedocs.io/en/latest/submit/assembly/genome.html#manifest-files. For Genbank submissions, instructions can be found here: https://www.ncbi.nlm.nih.gov/sra/docs/submit/. Submission of assemblies and raw reads with the above recommended Bioproject umbrella structure will ensure linking and discoverability.

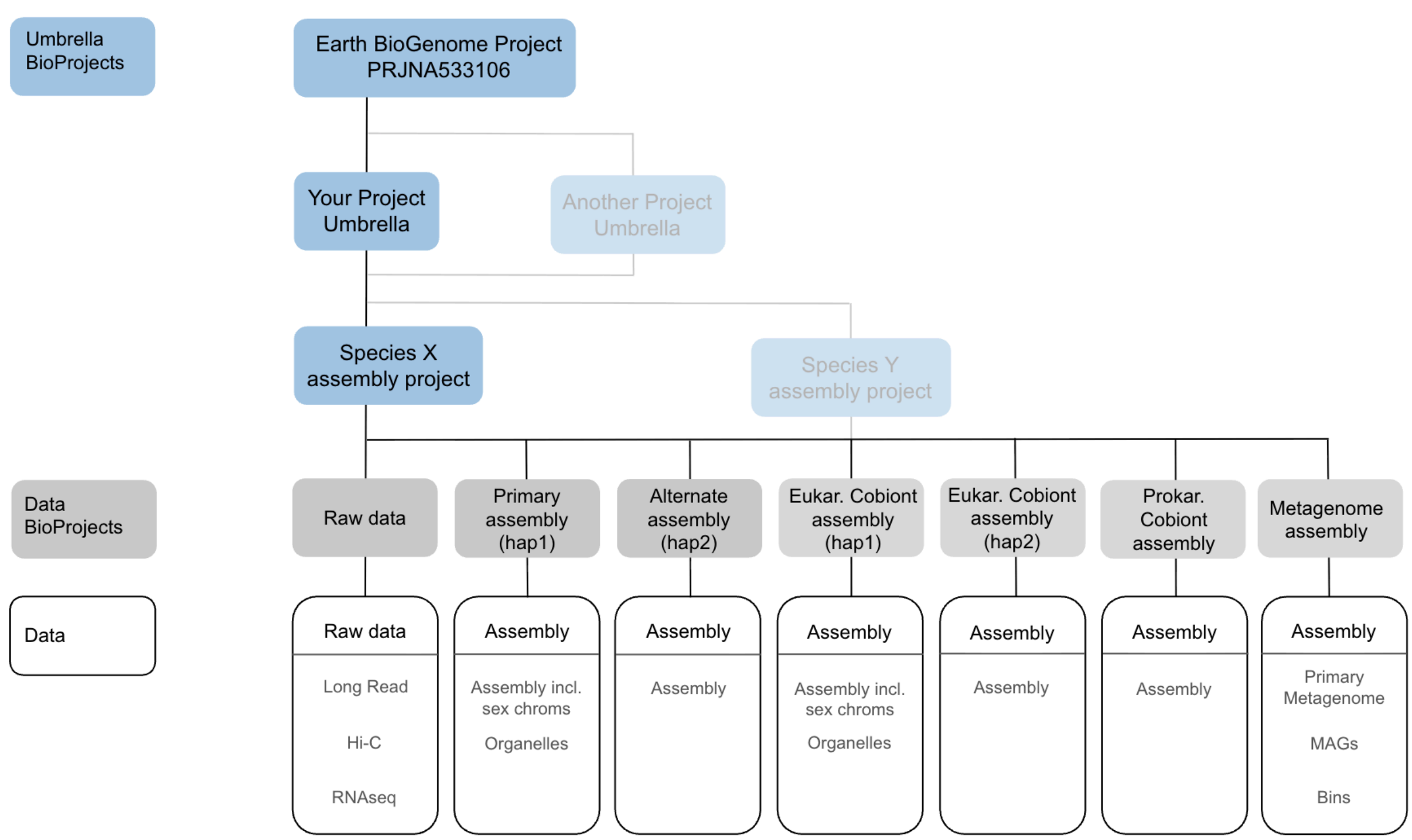

**Figure 2: Overview of the INSDC BioProject, assembly and raw data submission structure.** Add haplotype assembly BioProjects and data as required in case of polyploid eukaryotes. Equally, add as required for multiple prokaryotic assemblies, e.g. Wolbachias. Eukaryotic cobionts will require their own ToLID and BioProject. Umbrella projects can be linked and stacked to reflect their relationships. If several individuals of the same species are assembled and released by the same project, their Data BioProjects can be added to the species umbrella BioProject. This ensures the discoverability of all data for the relevant species, whilst the different ToLIDs for the individual specimens ensure identification of what data belongs together.

# Conclusion

We believe the standards specified above are an essential foundation for high-quality genome annotation and analysis supporting the scientific and societal goals of the EBP and the research community as a whole. However, we note the need for and encourage research to address the following problems and increase the proportion of species reaching this assembly standard: 1) Generating uniform coverage of long read sequences from low input DNA; 2) High quality phased assembly of high heterozygosity and/or polyploid genomes; and 3) Better genome sequencing and assembly of mixtures of unculturable single cell eukaryotes.

No specific sequencing recommendations are made, to allow for future technology improvement and change. We urge all genome projects to share their sequencing recipes and assembly pipelines via publication, protocols.io and github for reuse by others. Multiple long read plus Hi-C and/or other scaffolding strategies, for example the Vertebrate Genomes Project assembly pipeline (available on Galaxy, https://galaxyproject.org/projects/vgp/workflows/) or the Tree of Life assembly pipeline (https://github.com/sanger-tol/genomeassembly), now regularly reach these standards given sufficient material for less than the projected EBP phase 1 cost of $30,000 (Lewin et al. 2018). Indeed, we believe that it is now possible to reach these standards at substantially lower cost, of the order of $6,000 in direct costs for a 1 Gb genome. Because of this rapid change, the EBP sequencing and assembly committee is optimistic about reducing sequencing complexity and computational costs in the next few years to achieve the estimated direct cost for phases 2 and 3 of the EBP of $1900/species (Blaxter et al. 2025). We are aware that costs vary widely across different countries and that some regions even experienced significant price increases in the recent past (Vilaça et al. 2024). We recommend that cost reduction and the development of local capacity be treated as integral to, not separate from, the goal of generating universal high-quality reference genomes.

To monitor developments we will continue to revisit the standards on an annual basis, reconsidering what is feasible in terms of accuracy, completeness, and cost. It will be critical for the EBP to achieve its goal of generating high-quality reference genomes for all named eukaryotic species in the next 10 years that the global research community adopts these standards or even outperforms them.

# Declarations

## Acknowledgements

We gratefully acknowledge all members of the EBP Sequencing and Assembly Subcommittee and the participants of the Friday assembly calls (vgp-assembly googlegroup) for their valuable discussions and contributions, which informed the development and revision of the standards and recommendations summarised in this manuscript.

## Funding

MB, PD, CH, KH, MM, ZN, YS, CGSC, MUS are funded through Wellcome 220540/Z/20/A; JMA through ANR-22-EXAT-0003-SEQ-Sea; RMW by the Biodiversity Genomics Europe plus grant agreement ID: 101289347 (HORIZON-CL6-2025-01-BIODIV-03); MWB by the Gordon and Betty Moore Foundation (GBMF13832); TA by the Spanish Government, Ministry of Science, Innovation and Universities and Generalitat de Catalunya through the Departament de Recerca i Universitats and Departament de Salut; SK, and BJP by the Intramural Research Program of the National Institutes of Health (NIH); CY by the National Key R&D Program of China (2025YFC3410300); YZ by the National Natural Science Foundation of China (32400494); HL through NIH R01HG010040; GF and EDJ through NIH R03HG013362 and HHMI; JJS through NIH R35GM130349; PAFG through Sao Paulo Research Foundation #18/15579-8 and #25/18246-3; RD through Wellcome 317408/Z/24/Z and JO through ZIAHG000183-25.

## Data Availability

Not applicable

## Competing interests

SK received funds to speak at ONT events (prior to 2023). The other authors declare that they have no competing interests.

## Ethics approval and consent to participate

Not applicable

## Author contributions

CH, CY, EDJ, GF, GZ, HC, KH, MWB, RD, YZ and HL are members of the EBP sequencing and assembly standards subcommittee, developing the underlying standards and recommendations. RD led the construction of first versions of both these documents. KH coordinated their yearly updates, merged existing documents to create the draft manuscript, adapted it for publication, created figures and tables and consolidated all edits and commentary. All other authors provided edits and/or commentary. All authors have read and agreed with the content.

# Supplementary Material

**Supplementary Table 1**: Software tools for assembly processes

| Name | Purpose | Reference/Link |
|---|---|---|
| KMC | Kmer assessment | (Kokot et al. 2017) |
| FastK | Kmer assessment | https://github.com/thegenemyers/FASTK |
| GenomeScope 2.0 | Kmer assessment | (Ranallo-Benavidez et al. 2020) |
| SmudgePlot | Kmer assessment | (Ranallo-Benavidez et al. 2020) |
| Purge_dups | Duplication removal | (Guan et al. 2020) |
| BUSCO | Gene space completeness | (Tegenfeldt et al. 2025) |
| oatK | Organelle assembly | (Zhou et al. 2025) |
| MitoHiFi | Organelle assembly | (Uliano-Silva et al. 2023) |
| MitoVGP | Organelle assembly | (Formenti et al. 2021) |
| SALSA2 | Hi-C scaffolding | (Ghurye et al. 2019) |
| HiRise | Hi-C scaffolding | https://github.com/DovetailGenomics/HiRise_July2015_GR |
| 3D-DNA | Hi-C scaffolding | (Dudchenko et al. 2017) |
| wf-pore-c | CiFi/PoreC alignment | https://github.com/epi2me-labs/wf-pore-c |
| falign | CiFi/PoreC alignment | https://github.com/xiaochuanle/Falign |
| BlobToolkit | Cobiont visualisation | (Laetsch and Blaxter 2017) |
| FCS-GX | Cobiont separation | (Astashyn et al. 2024) |
| ASCC | Cobiont separation | (Pointon et al. 2025) |
| Kraken2 | Cobiont separation | (Wood et al. 2019) |
| hifiasm | Contig assembly | (Cheng et al. 2021) |
| metaMDBG | Metagenome assembly | (Benoit et al. 2024) |
| Myloasm | Metagenome assembly | (Shaw et al. 2026) |
| CheckM2 | Metagenome assessment | (Chklovski et al. 2023) |
| bwa | Short read alignment | (Li 2013) |
| FreeBayes | Variant calling | (Garrison and Marth 2012) |
| DeepVariant | Variant calling | (Poplin et al. 2018) |
| Merfin | Variant filtering | (Formenti et al. 2022) |
| bcftools | Sequence corrections | (Danecek et al. 2021) |
| WinnowMap | Long read mapping | (Jain et al. 2022) |
| PretextView | Manual curation | (Wood JMD, Harry E, Gilbert J, Guan S, Pointon D-L, Sims Y, Absolon DE, Brooks K, Collins J, Paulini M, Pelan S, Santos C, Zilov D, Ning Z, Mathers TC, Howe K, n.d.) |
| JuiceBox | Manual curation | (Durand et al. 2016) |
| MicroFinder | Microchromosome identification | (Mathers et al. 2026) |
| Verkko | T2T assembler for PacBio/ONT data | (Rautiainen et al. 2023) |
| SQK-APK114, SQK-LSK114, SQK-ULK114 | T2T assembler for ONT data | https://nanoporetech.com/document/telomere-to-telomere-sequencing-t2t-on-promethion-sqk-apk114-sqk |
| TTT | Manual T2T curation | (Antipov et al. 2026) |
| Verkko-fillet | Manual T2T curation | (Kim et al. 2026) |

**Supplementary Table 2:** Database for karyotypic and chromosome count data

| Group | Taxa | Resource | Link |
|---|---|---|---|
| Vertebrata | Fish | Fish Karyome | https://mail.nbfgr.res.in/Fish_Karyome/ |
| Vertebrata | Amphibia | Amphibian karyotype database | https://evobir.shinyapps.io/AmphibianDB/ |
| Vertebrata | Aves | Bird Chromosome Database | https://sites.unipampa.edu.br/birdchromosomedatabase/chromosome-painting-data/ |
| Hexapoda | Coleoptera | Coleoptera Karyotype Database | https://evobir.shinyapps.io/ColeopteraDB/ |
| Hexapoda | Diptera | Diptera Karyotype Database | https://evobir.shinyapps.io/Diptera/ |
| Hexapoda | Drosophila | Drosophoila Karyotype Database | https://evobir.shinyapps.io/Drosophila/ |
| Hexapoda | Polyneoptera | Polyneoptera Karyotype Database | https://evobir.shinyapps.io/PolyneopteraDB/ |
| Hexapoda | Blattodea | Blattodea Karyotype Database | http://web.natur.cuni.cz/zoologie/arthropods/blattodeadatabase/index.html |
| Hexapoda | Formicidae | Ant Chromosome Database | http://www.ants.ufop.br |
| Plants | Plants | Chromosome Counts Database | http://ccdb.tau.ac.il |
| Plants, Animals, Fungi | Plants, Animals, Fungi | B-chrom | http://www.bchrom.csic.es |
| Metazoa | Animals | Animal Chromosome Count Database | https://cromanpa94.github.io/ACC/category/other.html |
| Arthropoda | Animals | Arthropoda Cytogenetics | https://arthropodacytogenetics.bio.br/spiderdatabase/ |

# Supplementary References